\documentclass[12pt]{article}
\usepackage{graphicx}
\usepackage{epsf}
\def\dps{\displaystyle}

\def\beq{\begin{equation}}
\def\eeq{\end{equation}}
\def\be{\begin{equation}}
\def\ee{\end{equation}}
\def\ba{\begin{array}}
\def\ea{\end{array}}
\def\bt{\begin{tabular}}
\def\et{\end{tabular}}
\def\bea{\begin{eqnarray}}
\def\eea{\end{eqnarray}}
\def\beas{\begin{eqnarray*}}
\def\eeas{\end{eqnarray*}}
\def\bef{\begin{figure}}
\def\eef{\end{figure}}

\def\noi{\noindent}
\def\erm{{\mathrm e}}
\def\ul{\underline}
\def\bvec{\mathbf}
\def\ov{\overline}

\begin{document}


\title{\vspace{-4truecm} {}\hfill{\small DSF$-$06/2006} \\
\vspace{1truecm}
Four variations on Theoretical Physics \\ by Ettore Majorana}%
\author{S. Esposito \\
\footnotesize Dipartimento di Scienze Fisiche, Universit\`a di
Napoli ``Federico II'' \& \\
\footnotesize I.N.F.N. Sezione di Napoli, \\
\footnotesize Complesso Universitario di M. S. Angelo, Via
Cinthia,
80126 Naples, Italy \\
\small ({\rm Salvatore.Esposito@na.infn.it})}%


\date{}%

\maketitle

\begin{abstract}
\noi An account is given of some topical unpublished work by
Ettore Majorana, revealing his very deep intuitions and
skilfulness in Theoretical Physics. The relevance of the quite
unknown results obtained by him is pointed out as well.
\end{abstract}


\section{Introduction}

\noindent Probably, the highest appraisal received by the work of
Ettore Majorana was expressed by the Nobel Prize Enrico Fermi in
several occasions \cite{aaa}, but such opinions could appear as
overstatements or unjustified (especially because they are
expressed by a great physicist as Fermi), when compared with the
spare (known) Majorana's scientific production, just 9 published
papers. However, today the name of Majorana is largely known to
the nuclear and subnuclear physicist's community: Majorana
neutrino, Majorana-Heisenberg exchange forces, and so on are, in
fact, widely used concepts.

In this paper, we focus on the less-known (or completely unknown)
work by this scientist, aimed to shed some light on the peculiar
abilities of Majorana that were well recognized by Fermi and his
coworkers. The wide unpublished scientific production by Majorana
is testified by a large amount of papers \cite{catalog}, almost
all deposited at the Domus Galilaeana in Pisa; those known, in
Italian, as ``Volumetti" has been recently collected and
translated in a book \cite{volumetti}, and we refer the interested
reader to this book for further study.

Here we have chosen to discuss only four topics dealt with by
Majorana in different areas of Physics, just to give a sample of
his very deep intuitions and skilfulness, together with the
relevance of the results obtained.

We start with a discussion of a peculiar approach to Quantum
Mechanics, as deduced by a manuscript \cite{path} which probably
corresponds to the text for a seminar delivered at the University
of Naples in 1938, where Majorana lectured on Theoretical Physics
\cite{lezioni}. Some passages of that manuscript reveal a physical
interpretation of the Quantum Mechanics, which anticipates of
several years the Feynman approach in terms of path integrals,
independently of the underlying mathematical formulation. The main
topic of that dissertation was the application of Quantum
Mechanics to the theory of molecular bonding, but the present
scientific interest in it is more centered on the interpretation
given by Majorana about some topics of the novel, for that time,
Quantum Theory (namely, the concept of quantum state) and the
direct application of this theory to a particular case (that is,
precisely, the molecular bonding). It not only discloses a
peculiar cleverness of the author in treating a pivotal argument
of the novel Mechanics, but, keeping in mind that it was written
in 1938, also reveals a net advance of at least ten years in the
use made of that topic.

In the second topic, we report on a more applicative subject,
discussing an original method that leads to a semi-analytical
series solution of the Thomas–Fermi equation, with appropriate
boundary conditions, in terms of only one quadrature \cite{ajp}.
This was developed by Majorana in 1928, just when starting to
collaborate (still as a University student) with the Fermi group
in Rome, and reveals an outstanding ability to solve very involved
mathematical problems in a very interesting and clear way. The
whole work performed on the Thomas-Fermi model is contained in
some spare sheets, and diligently reported by the author himself
in his notebooks \cite{volumetti}. From these it is evident the
considerable contribution given by Majorana even in the
achievement of the statistical model \cite{DiGrezia},
anticipating, in many respects, some important results reached
later by leading specialists. But the major finding by Majorana
was his solution (or, rather, methods of solutions) of the
Thomas-Fermi equation, which remained completely unknown, until
recent times, to the Physics community, who ignored that the
non-linear differential equation relevant for atoms and other
systems could even be solved semi-analytically. The method
proposed by Majorana can also be extended to an entire class of
particular differential equations \cite{ijtp}.

Afterwards we discuss a subject that was repeatedly studied by
Majorana in his research notebooks; namely that of a formulation
of Electrodynamics in terms of the electric and magnetic fields,
rather than the potentials, which is suitable for a quantum
generalization, in a complete analogy with the Dirac theory
\cite{baldo} \cite{giannetto}. This argument was already faced in
1931 by Oppenheimer \cite{opp}, who only supposed the analogy of
the photon case with that described by Dirac, but Majorana
explicitly deduced a Dirac-like equation for the photon, thus
building up the presumed analogy.

Finally, we report on another topic particularly loved by
Majorana, after the appearance (at the end of 1928) of the seminal
book by Hermann Weyl \cite{Weyl28}, that is the Group Theory and
its application to physical problem. As testified by the large
number of unpublished manuscript pages of the Italian physicist,
the Weyl approach greatly influenced the scientific thought and
work of Majorana \cite{weylmajo}. In fact, when Majorana became
aware of the great relevance of the Weyl's application of the
Group Theory to Quantum Mechanics, he immediately grabbed the Weyl
method and developed it in many applications. In one of his
notebooks \cite{volumetti} we find, for example, a preliminary
study of what will be one of the most important (published) papers
by Majorana on a generalization of the Dirac equation to particles
with arbitrary spin \cite{infinite}. In particular, in 1932
Majorana obtained the infinite-dimensional unitary representations
of the Lorentz group that will be re-discovered by Wigner in his
1939 and 1948 works \cite{rediscover}, and the entire theory was
re-invented by Soviet mathematicians (in particular Gel'fand and
collaborators) in a series of articles from 1948 to 1958
\cite{Gelfand} and finally applied by physicists years later.

What presented here is, necessarily, a very short account of what
Majorana really did in his few years of work (about ten years),
but, we hope, it serves in the centennial year at least to
understand the very relevant role played by him in the advancement
of Physics.

\section{Path-Integral approach to Quantum Mechanics}

\noindent The usual quantum-mechanical description of a given
system is strongly centered on the role played by the hamiltonian
$H$ of the system and, as a consequence, the time variable plays
itself a key role in this description. Such a dissymmetry between
space and time variables is, obviously, not satisfactory in the
light of the postulates of the Theory of Relativity. This was
firstly realized in 1932 by Dirac \cite{dirac}, who put forward
the idea of reformulating the whole Quantum Mechanics in terms of
lagrangians rather than hamiltonians. The starting point in the
Dirac thought is that of exploiting an analogy, holding at the
quantum level, with the Hamilton principal function $S$ in
Classical Mechanics, thus writing the transition amplitude from
one space-time point to another as an (imaginary) exponential of
$S$. However, the original Dirac formulation was not free from
some unjustified assumptions, leading also to wrong results, and
the correct mathematical formulation and the physical
interpretation of it came only in the forties with the work by
Feynman \cite{feynman}. In practice, in the Feynman approach to
Quantum Mechanics, the transition amplitude between an initial and
a final state can be expressed as a sum of the factor
$\erm^{iS[q]/\hbar}$ over {\it all} the paths $q$ with fixed
end-points, not just those corresponding to classical dynamical
trajectories, for which the action is stationary.

In 1938 Majorana was appointed as full professor of Theoretical
Physics at the University of Naples, where probably delivered a
general conference mentioning also his particular viewpoint on
some basic concepts on Quantum Mechanics (see Ref. \cite{path}).
Fortunately enough, we have some papers written by him on this
subject, and few crucial points, anticipating the Feynman approach
to Quantum Mechanics, will be discussed in the following. However,
we firstly note that such papers contain {\it nothing} of the
mathematical aspect of that peculiar approach to Quantum
Mechanics, but it is quite evident as well the presence of the
{\it physical} foundations of it. This is particularly impressive
if we take into account that, in the known historical path, the
interpretation of the formalism has only followed the mathematical
development of the formalism itself.

The starting point in Majorana is to search for a meaningful and
clear formulation of the concept of quantum state. And, obviously,
in 1938 the dispute is opened with the conceptions of the Old
Quantum Theory.
\begin{quote}
According to the Heisenberg theory, a quantum state corresponds
not to a strangely privileged solution of the classical equations
but rather to a set of solutions which differ for the initial
conditions and even for the energy, i.e. what it is meant as
precisely defined energy for the quantum state corresponds to a
sort of average over the infinite classical orbits belonging to
that state. Thus the quantum states come to be the minimal
statistical sets of classical motions, \ul{slightly different}
from each other, accessible to the observations. These minimal
statistical sets cannot be further partitioned due to the
uncertainty principle, introduced by Heisenberg himself, which
forbids the precise simultaneous measurement of the position and
the velocity of a particle, that is the determination of its
orbit.
\end{quote}
Let us note that the ``solutions which differ for the initial
conditions" correspond, in the Feynman language of 1948, precisely
to the different integration paths. In fact, the different initial
conditions are, in any case, always referred to the same initial
time ($t_a$), while the determined quantum state corresponds to a
fixed end time ($t_b$). The introduced issue of ``\ul{slightly
different} classical motions" (the emphasis is given by Majorana
himself), according to what specified by the Heisenberg's
uncertainty principle and mentioned just afterwards, is thus
evidently related to that of the sufficiently wide integration
region required in the Feynman path-integral formula for quantum
(rather than classical) systems. In this respect, such a
mathematical point is intimately related to a fundamental physical
principle.

The crucial point in the Feynman formulation of Quantum Mechanics
is, as well-known, to consider not only the paths corresponding to
classical trajectories, but {\it all} the possible paths joining
the initial point with the end one. In the Majorana manuscript,
after a discussion on an interesting example on the harmonic
oscillator, the author points out:
\begin{quote}
Obviously the correspondence between quantum states and sets of
classical solutions is only approximate, since the equations
describing the quantum dynamics are in general independent of the
corresponding classical equations, but denote a real modification
of the mechanical laws, as well as a constraint on the feasibility
of a given observation; however it is better founded than the
representation of the quantum states in terms of quantized orbits,
and can be usefully employed in qualitative studies.
\end{quote}
And, in a later passage, it is more explicitly stated that the
wave function ``corresponds in Quantum Mechanics to any possible
state of the electron". Such a reference, that only superficially
could be interpreted, in the common acceptation, that all the
information on the physical systems is contained in the wave
function, should instead be considered in the meaning given by
Feynman, according to the comprehensive discussion made by
Majorana on the concept of state.

Finally we point out that, in the Majorana analysis, a key role is
played by the symmetry properties of the physical system.
\begin{quote}
Under given assumptions, that are verified in the very simple
problems which we will consider, we can say that every quantum
state possesses all the symmetry properties of the constraints of
the system.
\end{quote}
The relationship with the path-integral formulation is made as
follows. In discussing a given atomic system, Majorana points out
how from one quantum state $S$ of the system we can obtain another
one $S'$ by means of a symmetry operation.
\begin{quote}
However, differently from what happens in Classical Mechanics for
the \ul{single solutions} of the dynamical equations, in general
it is no longer true that $S'$ will be distinct from $S$. We can
realize this easily by representing $S'$ with a set of classical
solutions, as seen above; it then suffices that $S$ includes, for
any given solution, even the other one obtained from that solution
by applying a symmetry property of the motions of the systems, in
order that $S'$ results to be identical to $S$.
\end{quote}
This passage is particularly intriguing if we observe that the
issue of the redundant counting in the integration measure in
gauge theories, leading to infinite expressions for the transition
amplitudes, was raised (and solved) only after much time from the
Feynman paper.

\section{Solution of the Thomas-Fermi equation}

\noindent The main idea of the Thomas-Fermi atomic model is that
of considering the electrons around the nucleus as a gas of
particles, obeying the Pauli exclusion principle, at the absolute
zero of temperature. The limiting case of the Fermi statistics for
strong degeneracy applies to such a gas. Then, in this
approximation, the potential $V$ inside a given atom of charge
number $Z$ at a distance $r$ from the nucleus may be written as %
\begin{equation}
V(r) = \frac{Ze}{r}\varphi (r) . \label{9}
\end{equation}
With a suitable change of variable, $r = \mu x$ and
\begin{equation}
\mu =
\frac{1}{2}\left(\frac{3\pi}{4}\right)^{2/3}\frac{\hbar^2}{m_e
e^2}Z^{-1/3}  , \label{10}
\end{equation}
the Thomas-Fermi function $\varphi$ satisfies the following
non-linear differential equation (for $\varphi > 0$):
\begin{equation}
\varphi^{\prime\prime} = \frac{\varphi^{3/2}}{\sqrt{x}} \label{11}
\end{equation}
(a prime denotes differentiation with respect to $x$) with the
boundary conditions:
\begin{equation}
\begin{array}{rcl}
\varphi (0) & = & 1,
\\ & & \\
\varphi (\infty) & = & 0.
\end{array}
\label{12}
\end{equation}
The Fermi equation (\ref{11}) is a universal equation which does
not depend neither on $Z$ nor on physical constants ($\hbar, m ,
e$). Its solution gives, from Eq. (\ref{9}), as noted by Fermi
himself, a screened Coulomb potential which at any point is equal
to that produced by an effective charge
\begin{equation}
Z e \, \varphi\left(\frac{r}{\mu}\right). \label{13}
\end{equation}
As was immediately realized, in force of the independence of Eq.
(\ref{11}) on $Z$, the method gives an effective potential which
can be easily adapted to describe any atom with a suitable scaling
factor, according to Eq. (\ref{13}).

The problem of the theoretical calculation of observable atomic
properties is thus solved, in the Thomas-Fermi approximation, in
terms of the function $\varphi (x)$ introduced in Eq. (\ref{9})
and satisfying the Fermi differential equation (\ref{11}). By
using standard but involved mathematical tools, in his paper
\cite{[Thomas]} Thomas got an exact, ``singular" solution of his
differential equation satisfying only the second condition
(\ref{12}). This was later (in 1930) considered by Sommerfeld
\cite{[Sommerfeld]} as an approximation of the function $\varphi
(x)$ for large $x$ (and is indeed known as the ``Sommerfeld
solution" of the Fermi equation),
\begin{equation}
\varphi (x)=   \frac{144}{x^3} , \label{16}
\end{equation}
and Sommerfeld himself obtained corrections to the above quantity
in order to approximate in a better way the function $\varphi (x)$
for not extremely large values of $x$.

Until recent times it has been believed that the solution of such
equation satisfying both the appropriate boundary conditions in
(\ref{12}) cannot be expressed in closed form, and some effort has
been made, starting from Thomas \cite{[Thomas]}, Fermi
\cite{[FN43]}, \cite{[FN49]} and others, in order to achieve the
numerical integration of the differential equation. However, we
now know \cite{ajp}, \cite{DiGrezia} that Majorana in 1927-8 found
a semi-analytical solution of the Thomas-Fermi equation by
applying a novel exact method \cite{ijtp}. Before proceeding, we
will indulge here on an anecdote reported by Rasetti \cite{[FNM]},
Segr\`{e} \cite{[Segre]} and Amaldi \cite{[Amaldi]}. According to
the last author, ``Fermi gave a broad outline of the model and
showed some reprints of his recent works on the subject to
Majorana, in particular the table showing the numerical values of
the so-called Fermi universal potential. Majorana listened with
interest and, after having asked for some explanations, left
without giving any indication of his thoughts or intentions. The
next day, towards the end of the morning, he again came into
Fermi's office and asked him without more ado to draw him the
table which he had seen for few moments the day before. Holding
this table in his hand, he took from his pocket a piece of paper
on which he had worked out a similar table at home in the last
twenty-four hours, transforming, as far as Segr\`{e} remembers,
the second-order Thomas-Fermi non-linear differential equation
into a Riccati equation, which he had then integrated
numerically."

The whole work performed by Majorana on the solution of the Fermi
equation, is contained in some spare sheets conserved at the Domus
Galilaeana in Pisa, and diligently reported by the author himself
in his notebooks \cite{volumetti}. The reduction of the Fermi
equation to an Abel equation (rather than a Riccati one, as
confused by Segr\`{e}) proceeds as follows. Let's adopt a change
of variables, from $(x, \varphi)$ to $(t, u)$, where the formula
relating the two sets of variables has to be determined in order
to satisfy, if possible, both the boundary conditions (\ref{12}).
The function $\varphi$ in Eq. (\ref{16}) has the correct behavior
for large $x$, but the wrong one near $x=0$, so that we could
modify the functional form of $\varphi$ to take into account the
first condition in (\ref{12}). An obvious modification is
$\varphi=(144/x^3)f(x)$, with $f(x)$ a suitable function which
vanishes for $x\rightarrow 0$ in order to account for $\varphi
(x=0) = 1$. The simplest choice for $f(x)$ is a polynomial in the
novel variable $t$, as it was also considered later, in a similar
way, by Sommerfeld \cite{[Sommerfeld]}. The Majorana choice is as
follows:
\begin{equation}
\varphi (x)= \frac{144}{x^3}(1-t)^2 , \label{23}
\end{equation}
with $t\rightarrow 1$ as $ x\rightarrow 0$. From Eq. (\ref{23}) we
can then obtain the first relation linking $t$ to $x, \varphi$.
The second one, involving the dependent variable $u$, is that
typical of homogeneous differential equations (like the Fermi
equation) for reducing the order of the equation, i.e.
exponentiation with an integral of $u(t)$. The transformation
relations are thus:
\begin{equation}
\begin{array}{rcl}
t &=& \displaystyle 1- \frac{1}{12}\sqrt{x^3\varphi},
\\ & & \\
\varphi &=& e^{\int_1^t u(t) dt}\, .
\end{array}
\label{24}
\end{equation}
Substitution into Eq. (\ref{11}) leads to an Abel equation for
$u(t)$,
\begin{equation}
\frac{du}{dt}= \alpha (t) + \beta (t)\, u + \gamma (t)\, u^2 +
\delta (t) \, u^3 , \label{25}
\end{equation}
with
\begin{equation}
\begin{array}{rcl}
\alpha (t) &=& \displaystyle \frac{16}{3(1-t)},
\\ & & \\
\beta (t) &=& \displaystyle 8 +  \frac{1}{3(1-t)},
\\ & & \\
\gamma (t) &=& \displaystyle \frac{7}{3} - 4t,
\\ & & \\
\delta (t) &=& \displaystyle -\frac{2}{3}t(1 - t) .
\end{array}
\label{26}
\end{equation}
Note that both the boundary conditions in (\ref{12}) are
automatically verified by the relations (\ref{24}). We have
reported the derivation of the Abel equation (\ref{25}) mainly for
historical reasons (nevertheless, it is quite important since, in
this way, all the theorems on the Abel equation may thus be
applied to the non-linear Thomas-Fermi equation too); the precise
numerical values for the Fermi function $\varphi (x)$ were
obtained by Majorana by solving a different first-order
differential equation. Instead of Eq. (\ref{23}), Majorana chooses
$\varphi (x) $ of the form
\begin{equation}
\varphi (x)= \frac{144}{x^3} \, t^{6} . \label{27}
\end{equation}
Now the point with $x=0$ corresponds to $t=0$. In order to obtain
again a first order differential equation for $u(t)$, the
transformation equation for the variable $u$ involves $\varphi$
and its first derivative. Majorana then introduced the following
formulas:
\begin{equation}
\begin{array}{rcl}
t &=& 144^{-1/6} \, x^{1/2} \, \varphi^{1/6}, \nonumber \\ & & \\
u &=& \displaystyle -\left(\frac{16}{3}\right)^{1/3}
\varphi^{-4/3} \varphi^\prime \, \, .
\end{array}
\label{28}
\end{equation}
By taking the $t$-derivative of the last equation in (\ref{28})
and inserting Eq. (\ref{11}) in it, one gets:
\begin{equation}
\frac{du}{dt}= -
\left(\frac{16}{3}\right)^{1/3}\dot{x}\varphi^{-4/3}\left[-\frac{4}{3}
\frac{\varphi^{\prime 2}}{\varphi} +
\frac{\varphi^{3/2}}{x^{1/2}}\right] . \label{29}
\end{equation}
By using Eqs. (\ref{28}) to eliminate $x^{1/2}$ and
$\varphi^{\prime 2}$, the following equation results:
\begin{equation}
\frac{du}{dt}= \left(\frac{4}{9}\right)^{1/3}\frac{tu^2
-1}{t}\dot{x}\varphi^{1/3} . \label{30}
\end{equation}
Now the quantity $\dot{x}\varphi^{1/3}$ can be expressed in terms
of $t$ and $u$ by making use again of the first equation in
(\ref{28}) (and its $t$-derivative). After some algebra, the final
result for the differential equation for $u(t)$ is:
\begin{equation}
\frac{du}{dt}= 8 \, \frac{tu^2 -1}{1-t^2u} . \label{31}
\end{equation}
The obtained equation is again non-linear but, differently from
the original Fermi equation (\ref{11}), it is first-order in the
novel variable $t$ and the degree of non-linearity is lower than
that of Eq. (\ref{11}). The boundary conditions for $u(t)$ are
easily taken into account from the second equation in (\ref{28})
and by requiring that for $x\rightarrow\infty$ the Sommerfeld
solution (Eq. (\ref{27}) with $t=1$) be recovered:
\begin{equation}
\begin{array}{rcl}
u(0) &=& \displaystyle - \left(\frac{16}{3}\right)^{1/3}
\varphi^\prime_0 ,
\\ & & \\
u(1) &=&  1 .
\end{array}
\label{32}
\end{equation}
Here we have denoted with $\varphi^\prime _0 =
\varphi^\prime(x=0)$ the initial slope of the Thomas-Fermi
function $\varphi (x)$ which, for a neutral atom, is approximately
equal to $-1.588$.

The solution of Eq. (\ref{31}) was achieved by Majorana in terms
of a series expansion in powers of the variable $\tau = 1-t$:
\begin{equation}
u= a_0 + a_1 \tau + a_2 \tau^2 + a_3 \tau^3 + ...\,\, . \label{33}
\end{equation}
Substitution of Eq. (\ref{33}) (with the conditions in Eq.
(\ref{32})) into Eq. (\ref{31}) results into an iterative formula
for the coefficients $a_n$ (for details see Ref. \cite{ajp}). It
is remarkable that the series expansion in Eq. (\ref{33}) is
uniformly convergent in the interval $[0,1]$ for $\tau$, since the
series $\sum_{n=0}^\infty a_n$ of the coefficients converges to a
finite value determined by the initial slope $\varphi^\prime_0$.
In fact, by setting $\tau = 1$ ($t=0$) in Eq. (\ref{33}) we have
from Eq. (\ref{32}):
\begin{equation}
\sum_{n=0}^\infty a_n =  -
\left(\frac{16}{3}\right)\varphi^\prime_0 \label{34}
\end{equation}
Majorana was aware \cite{volumetti} of the fact that the series in
Eq. (\ref{33}) exhibits geometrical convergence with $a_n/
a_{n-1}\sim 4/5$ for $n\rightarrow\infty$.
\begin{figure}
\begin{center}
\epsfysize=5cm%
\epsffile{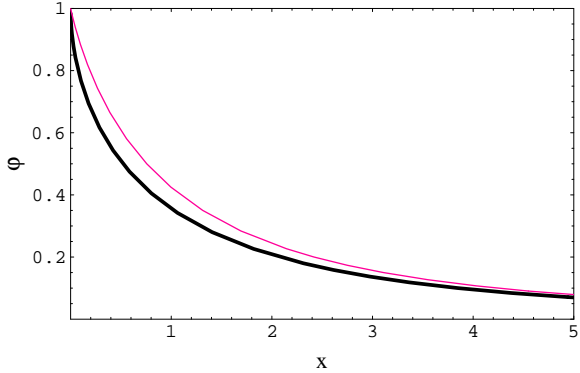}%
\caption{The Thomas-Fermi function $\varphi (x)$ and the Majorana
approximation of it. The thin (upper) line refers to the exact
(numerical) solution of Eq. (\ref{11}) while the thick (lower) one
corresponds to the parametric solution obtained from Eqs.
(\ref{36})-(\ref{37}).} \label{fig2}
\end{center}
\end{figure}

Given the function $u(t)$, we now have to look for the
Thomas-Fermi function $\varphi(x)$. This was obtained in a
parametric form $x=x(t)$, $\varphi = \varphi (t)$ in terms of the
parameter $t$ already introduced in Eq. (\ref{28}), and by writing
$\varphi (t)$ as
\begin{equation}
\varphi (t) = e^{\int_0^t w(t) dt} \label{35}
\end{equation}
(with this choice, $\varphi (t=0) =1$ and the first condition in
(\ref{12}) is automatically satisfied). Here $w(t)$ is an
auxiliary function which is determined in terms of $u(t)$ by
substituting Eq. (\ref{35}) into Eq. (\ref{28}). As a result, the
parametric solution of Eq. (\ref{11}), with boundary conditions
(\ref{12}), takes the form:
\begin{equation}
\begin{array}{rcl}
x(t)& = & 144^{1/3} \, t^2 \, e^{2{\cal I}(t)}
\\ & & \\
\varphi (t)& = & e^{-6 \, {\cal I}(t)}
\end{array}
\label{36}
\end{equation}
with
\begin{equation}
{\cal I}(t) = \int_0^t \frac{ut}{1-t^2u} \, dt \label{37}
\end{equation}
Remarkably, the Majorana solution of the Thomas-Fermi equation is
obtained with only one quadrature and gives easily obtainable
numerical values for the electrostatic potential inside atoms. By
taking into account only $10$ terms in the series expansion for
$u(t)$, such numerical values approximate the values of the exact
solution of the Thomas-Fermi equation with a relative error of the
order of $0.1\% $.

The intriguing property in the Majorana derivation of the solution
of the Thomas-Fermi equation is that his method can be easily
generalized and may be applied to a large class of particular
differential equations, as discussed in \cite{ijtp}.

Several generalizations of the Thomas-Fermi method for atoms were
proposed as early as in $1928$ by Majorana, but they were
considered by the physics community, ignoring the Majorana
unpublished works, only many years later.

Indeed, in Sect. 16 of Volumetto II \cite{volumetti}, Majorana
studied the problem of an atom in a weak external electric field
$E$, i.e. atomic polarizability, and obtained an expression for
the electric dipole moment for a (neutral or arbitrarily ionized)
atom.

Furthermore, he also started to consider the application of the
statistical method to molecules, rather than single atoms,
studying the case of a diatomic molecule with identical nuclei
(see Sect. 12 of Volumetto II \cite{volumetti}). The effective
potential in the molecule was cast in the form:

\begin{equation}
V= V_1 +V_2 -\alpha \, \frac{2V_1V_2}{V_1 + V_2}, \label{41}
\end{equation}

\noindent $V_1$ and $V_2$ being the potentials generated by each
of the two atoms. The function $\alpha$ must obey the differential
equation for $V$,
\begin{equation}
\nabla^2 V =-kV^{3/2} \label{42}
\end{equation}
($k$ is a suitable constant), with appropriate boundary
conditions, discussed in \cite{volumetti}. Majorana also gave a
general method to determine $V$ when the equipotential surfaces
are approximately known (see Sect. 12 of Volumetto III
\cite{volumetti}). In fact, writing the approximate expression for
the equipotential surfaces, as functions of a parameter $p$, as
\begin{equation}
f(x,y,z) =p, \label{43}
\end{equation}
he deduced a thorough equation from which it is possible to
determine $V(\rho)$, when the boundary conditions are assigned.
The particular case of a diatomic molecule with identical nuclei
was, again, considered by Majorana using elliptic coordinates in
order to illustrate his original method \cite{volumetti}.

Finally, our author also considered the second approximation for
the potential inside the atom, beyond the Thomas-Fermi one, with a
generalization of the statistical model of neutral atoms to those
ionized $n$ times, including the case $n = 0$ (see Sect. 15 of
Volumetto II \cite{volumetti}). As recently pointed out, the
approach used by Majorana to this end is rather similar to that
now adopted in the renormalization of physical quantities in
modern gauge theories \cite{comment}.

\section{Majorana formulation of Electrodynamics}

\noindent In 1931, in his ``note on light quanta and the
electromagnetic field" \cite{opp}, Oppenheimer developed an
alternative model to the theory of Quantum Electrodynamics,
starting from an analogy with the Dirac theory of the electron.
Such a formulation was particularly held dear by Majorana, who
studied it in some of his unpublished notebooks \cite{baldo}.

Majorana's original idea was that if the Maxwell theory of
electromagnetism has to be viewed as the wave mechanics of the
photon, then it must be possible to write the Maxwell equations as
a Dirac-like equation for a probability quantum wave $\psi$, this
wave function being expressible by means of the physical ${\cal
{\vec E}}$, ${\cal {\vec B}}$ fields. This can be, indeed,
realized introducing the quantity %
\be \vec{\psi} \; = \; {\cal
{\vec E}} \; - \; i \, {\cal {\vec B}} \label{m11} \ee%
since $\vec{\psi}^{\ast} {\cdot} \vec{\psi} \, = \, {\cal {\vec
E}}^2 \, + \, {\cal {\vec B}}^2$ is directly proportional to the
probability density function for a photon \footnote{If we have a
beam of $n$ equal photons each of them with energy $\epsilon$
(given by the Planck relation), since $\frac{1}{2} \, ({\cal {\vec
E}}^2 \, + \, {\cal {\vec B}}^2)$ is the energy density of the
electromagnetic field, then $\frac{1}{n \epsilon} \, \frac{1}{2}
\, ({\cal {\vec E}}^2 \, + \, {\cal {\vec B}}^2) \, dS \, dt$
gives the probability that each photon has to be detected in the
area $dS$ in the time $dt$. The generalization to photons of
different energies (i.e. of different frequencies) is obtained
with the aid of the superposition principle.}. In terms of
$\vec{\psi}$, the Maxwell equations in vacuum then write %
\bea \left. \right. & \left. \right. & \vec{\nabla} {\cdot}
\vec{\psi} \; =
\; 0 \label{m12} \\
\left. \right. & \left. \right. & \frac{\partial
\vec{\psi}}{\partial t} \; = \; i \, \vec{\nabla} {\times}
\vec{\psi} \label{m13} \eea %
By making use of the correspondence principle %
\bea
E & \rightarrow & i \, \frac{\partial}{\partial t} \label{14} \\
\vec{p} & \rightarrow & - i \, \vec{\nabla} \label{15} \eea %
these equations effectively can be cast in a Dirac-like form %
\be \left( E \; - \; \alpha {\cdot} \vec{p} \right) \, \vec{\psi}
\; = \; 0 \label{m16} \ee %
with the transversality condition %
\be \vec{p} {\cdot} \vec{\psi} \; = \; 0 \label{17} \ee %
where the 3x3 hermitian matrices $( \alpha_i )_{lm} \, = \, i \,
\epsilon_{i l m}$ %
\be \alpha^1 \; = \; \left( \ba{ccc}
0 & 0 & 0 \\
0 & 0 & i \\
0 & -i & 0 \ea \right) \;\;\;\;\;\;\;\;\;\; \alpha^2 \; = \;
\left( \ba{ccc}
0 & 0 & -i \\
0 & 0 & 0 \\
i & 0 & 0 \ea \right) \;\;\;\;\;\;\;\;\;\; \alpha^3 \; = \; \left(
\ba{ccc}
0 & i & 0 \\
-i & 0 & 0 \\
0 & 0 & 0 \ea \right) \label{18} \ee satisfying \be \left[
\alpha_i \; , \; \alpha_j \right] \; = \; - i \, \epsilon_{i j k}
\, \alpha_k \label{19} \\
\ee %
have been introduced.

The probabilistic interpretation is indeed possible
given the ``continuity equation'' (Poynting theorem) %
\be \frac{\partial \rho}{\partial t} \; + \; \vec{\nabla} {\cdot}
\vec{j} \; = \; 0 \label{110} \ee %
where %
\be \rho \; = \; \frac{1}{2} \, \vec{\psi}^{\ast} {\cdot}
\vec{\psi} \;\;\;\;\;\;\;\;\;\;\;\;\;\;\;\; \vec{j} \; = \; - \,
\frac{1}{2} \, \psi^{\ast} \, \alpha \, \psi \label{111} \ee %
are respectively the energy and momentum density of the
electromagnetic field.

It is interesting to observe that, differently from Oppenheimer,
who started from a mere, presumed analogy with the electron case,
Majorana built on analytically the analogy with the Dirac theory,
at a dynamical level, by {\it deducing} the Dirac-like equation
for the photon from the Maxwell equations with the introduction of
a complex wave field. As noted by Giannetto in Ref.
\cite{giannetto}, the Majorana formulation is algebraically
equivalent to the standard one of Quantum Electrodynamics and, in
addition, also some relevant problems concerning the negative
energy states, that induced Oppenheimer to abandon his model, may
be elegantly solved by using the method envisaged in a later work
\cite{elpos}, thus giving further physical insight into Majorana
theory.

\section{Lorentz group and its applications}

\noindent The important role of symmetries in Quantum Mechanics
was established in the third decade of the XX century, when it was
discovered the special relationships concerning systems of
identical particles, reflection and rotational symmetry or
translation invariance. Very soon it was discovered that the
systematic theory of symmetry resulted to be just a part of the
mathematical theory of groups, as pointed out, for example, in the
reference book by H. Weil \cite{Weyl28}). A particularly
intriguing example is that of the Lorentz group which, as well
known, underlies the Theory of Relativity, and its representations
are especially relevant for the Dirac equation in Relativistic
Quantum Mechanics. In the mentioned book, however, although the
correspondence between the Dirac equation and the Lorentz
transformations is pointed out, the group properties of this
connection are not highlighted. Moreover, only a particular kind
of such representations are considered (those related to the
two-dimensional representations of the group of rotations,
according to Pauli), but an exhaustive study of this subject was
still lacking at that time.

The situation changes \cite{weylmajo} quite sensibly with several
(unpublished) papers by Majorana \cite{volumetti}, where he gives
a detailed deduction of the relationship between the
representations of the Lorentz group and the matrices of the
(special) unitary group in two dimensions, and a strict connection
with the Dirac equation is always taken into account. Moreover the
{\it explicit} form of the transformations of every bilinear in
the spinor field $\Psi$ is reported. For example, Majorana obtains
that some of such bilinears behave as the 4-position vector
$(ct,x,y,z)$ or as the components of the rank-2 electromagnetic
tensor $({\bvec{E}},{\bvec{H}})$ under Lorentz transformations,
according to the following rules:
 \begin{eqnarray*}
& \Psi^\dagger \Psi \sim - i \Psi^\dagger \alpha_x \alpha_y
\alpha_z \Psi \sim c t, &  \\
& - \Psi^\dagger \alpha_x \Psi \sim i \Psi^\dagger
\alpha_y \alpha_z \Psi \sim x, &  \\
& - \Psi^\dagger \alpha_y \Psi \sim i \Psi^\dagger \alpha_z
\alpha_x \Psi \sim y, &   \\
& - \Psi^\dagger \alpha_z \Psi \sim i
\Psi^\dagger \alpha_x \alpha_y \Psi \sim z, & \\
& i \Psi^\dagger  \beta \alpha_x \Psi  \sim  E_x,\quad
 i \Psi^\dagger  \beta \alpha_y \Psi  \sim  E_y, \quad
 i \Psi^\dagger  \beta \alpha_z \Psi  \sim  E_z,  &   \\
& i \Psi^\dagger  \beta \alpha_y \alpha_z \Psi \sim  H_x ,\quad
 i \Psi^\dagger  \beta \alpha_z \alpha_x \Psi \sim  H_y ,\quad
 i \Psi^\dagger  \beta \alpha_x \alpha_y \Psi \sim  H_z, &  \\
& \Psi^\dagger  \beta \Psi \; \sim \; \Psi^\dagger  \beta \alpha_x
\alpha_y \alpha_z \Psi \; \sim \; 1, &
 \end{eqnarray*}
where $\alpha_x, \alpha_y, \alpha_z, \beta$ are Dirac matrices.

But, probably, the most important result achieved by Majorana on
this subject is his discussion of {\bf infinite-dimensional}
unitary representations of the Lorentz group, giving also an {\it
explicit} form for them. Note that such representations were
independently discovered by Wigner in 1939 and 1948
\cite{rediscover} and were thoroughly studied only in the years
1948-1958 \cite{Gelfand}. Lucky enough, we are able to reconstruct
the reasoning which led Majorana to discuss the
infinite-dimensional representations. In Sec. 8  of Volumetto V we
read \cite{volumetti}:
\begin{quote}
``The representations of the Lorentz group are, except for the
identity representation, essentially not unitary, i.e., they
cannot be converted into unitary representations by some
transformation. The reason for this is that the Lorentz group is
an open group. However, in contrast to what happens for closed
groups, open groups may have irreducible representations (even
unitary) in infinite dimensions. In what follows, we shall give
two classes of such representations for the Lorentz group, each of
them composed of a continuous infinity of unitary
representations.''
\end{quote}
The two classes of representations correspond to integer and
half-integer values for the representation index $j$ (angular
momentum). Majorana begins by noting that the group of the real
Lorentz transformations acting on the variables $ct, x, y, z$ can
be constructed from the infinitesimal transformations associated
to the matrices:
\begin{equation}
\begin{array}{c}
${}$ \\
\displaystyle S_x \; = \; \left( \ba{cccc}
                 0 & 0 & 0 & 0  \\
                 0 & 0 & 0 & 0  \\
                 0 & 0 & 0 & -1 \\
                 0 & 0 & 1 & 0
             \ea \right) ,\quad
S_y \; = \; \left(  \ba{cccc}
                 0 & 0 & 0 & 0 \\
                 0 & 0 & 0 & 1 \\
                 0 & 0 & 0 & 0 \\
                 0 & -1& 0 & 0
             \ea \right),
\\ ${}$ \\ \displaystyle
S_z \; = \; \left(  \ba{cccc}
                 0 & 0 & 0 & 0 \\
                 0 & 0 &-1 & 0 \\
                 0 & 1 & 0 & 0 \\
                 0 & 0 & 0 & 0
             \ea \right),
\\ ${}$ \\ \displaystyle
T_x \; = \; \left( \ba{cccc}
                 0 & 1 & 0 & 0 \\
                 1 & 0 & 0 & 0 \\
                 0 & 0 & 0 & 0 \\
                 0 & 0 & 0 & 0
             \ea \right) ,\quad
T_y \; = \; \left(  \ba{cccc}
                 0 & 0 & 1 & 0 \\
                 0 & 0 & 0 & 0 \\
                 1 & 0 & 0 & 0 \\
                 0 & 0 & 0 & 0
             \ea \right) ,
\\ ${}$ \\ \displaystyle
T_z \; = \; \left(  \ba{cccc}
                 0 & 0 & 0 & 1 \\
                 0 & 0 & 0 & 0 \\
                 0 & 0 & 0 & 0 \\
                 1 & 0 & 0 & 0
             \ea \right),
\end{array}
 \end{equation}
from which he deduces the general commutation relations satisfied
by the $S$ and $T$ operators acting on generic (even infinite)
tensors or spinors:
 \bea
S_x \, S_y \, - \, S_y \, S_x &=&  S_z ,\nonumber \\ T_x \, T_y \,
- \, T_y \, T_x &=& - \, S_z, \nonumber \\ S_x \, T_x \, - \, T_x
\, S_x &=&
0, \label{4.3a} \\ S_x \, T_y \, - \, T_y \, S_x &=& T_z, \nonumber \\
S_x \, T_z \, - \, T_z \, S_x &=& - \, T_y, \nonumber \\
{\rm{etc}}. \nonumber
 \eea
Next he introduces the matrices
 \be \label{4.4}
a_x \; = \; i \, S_x,\quad b_x \; = \; - i \, T_x,\quad
{\rm{etc.}}
 \ee
which are Hermitian for unitary representations (and viceversa),
and obey the following commutation relations:
 \bea
\left[ a_x ,~ a_y \right] &=& i \, a_z, \nonumber \\
\left[ b_x ,~ b_y \right] &=& - i \, a_z, \nonumber \\
\left[ a_x ,~ b_x \right] &=& 0, \label{4.4a} \\
\left[ a_x ,~ b_y \right] &=& i \, b_z,  \nonumber \\
\left[ b_x ,~ a_y \right] &=& i \, b_z,  \nonumber \\
 {\rm{etc}}. \nonumber
 \eea
By using only these relations he then obtains (algebraically
\footnote{The algebraic method to obtain the matrix elements in
Eq. (\ref{4.4b}) follows closely the analogous one for evaluating
eigenvalues and normalization factors for angular momentum
operators, discovered by Born, Heisenberg and Jordan in 1926 and
reported in every textbook on Quantum Mechanics (see, for example,
\cite{Sakurai}).}) the explicit expressions of the matrix elements
for given $j$ and $m$ \cite{volumetti} \cite{infinite}. The
non-zero elements of the infinite matrices $a$ and $b$, whose
diagonal elements are labelled by $j$ and $m$, are as follows:
 \bea
\!\!\!\!\! \dps
<j,m \, | \, a_x - i a_y \, | \, j,m+1> &=& \dps \sqrt{(j+m+1)(j-m)}, \nonumber \\
\!\!\!\!\! \dps
<j,m \, | \, a_x + i a_y \, | \, j,m-1> &=& \dps \sqrt{(j+m)(j-m+1)}, \nonumber \\
\!\!\!\!\! \dps
<j,m \, | \, a_z \, | \, j,m> &=& \dps m, \nonumber \\
\!\!\!\!\! \dps <j,m \, | \, b_x - i b_y \, | \, j+1,m+1> &=& \dps
- \, \frac{1}{2} \,
\sqrt{(j+m+1)(j+m+2)}, \nonumber \\
\!\!\!\!\! \dps <j,m \, | \, b_x - i b_y \, | \, j-1,m+1> &=& \dps
\frac{1}{2} \,
\sqrt{(j-m)(j-m-1)}, \label{4.4b} \\
\!\!\!\!\! \dps <j,m \, | \, b_x + i b_y \, | \, j+1,m-1> &=& \dps
\frac{1}{2} \,
\sqrt{(j-m+1)(j-m+2)}, \nonumber \\
\!\!\!\!\! \dps <j,m \, | \, b_x + i b_y \, | \, j-1,m-1> &=& \dps
- \, \frac{1}{2} \,
\sqrt{(j+m)(j+m-1)},  \nonumber \\
\!\!\!\!\! \dps <j,m \, | \, b_z \, | \, j+1,m> &=& \dps
\frac{1}{2} \, \sqrt{(j+m+1)(j-m+1)}, \nonumber \\
\!\!\!\!\! \dps <j,m \, | \, b_z \, | \, j-1,m> &=& \dps
\frac{1}{2} \, \sqrt{(j+m)(j-m)}. \nonumber
 \eea
The quantities on which $a$ and $b$ act are infinite tensors or
spinors (for integer or half-integer $j$, respectively) in the
given representation, so that Majorana effectively constructs, for
the first time, infinite-dimensional representations of the
Lorentz group. In \cite{infinite} the author also picks out a
physical realization for the matrices $a$ and $b$ for Dirac
particles with energy operator $H$, momentum operator $\bvec{p}$
and spin operator $\bvec{\sigma}$:
\begin{equation} \label{4.5}
 a_x \, = \, \frac{1}{\hbar} \left( y p_z - z p_y \right) +
 \frac{1}{2} \sigma_x , \quad b_x \, = \, \frac{1}{\hbar} x
 \frac{H}{c} + \frac{i}{2} \alpha_x , \quad {\rm{etc.}} ,
\end{equation}
where $\alpha_x, \alpha_y,\alpha_z$ are the Dirac
$\alpha$-matrices.

Further development of this material then brought Majorana to
obtain a relativistic equation for a wave-function $\psi$ with
infinite components, able to describe particles with arbitrary
spin (the result was published in 1932 \cite{infinite}). By
starting from the following variational principle:
\begin{equation}\label{4.5a}
  \delta \int \ov{\psi} \left( \frac{H}{c} + {\bvec{\alpha}} \cdot
  {\bvec{p}} - \beta m c \right) \psi \, d^4 x \, = \, 0 ,
\end{equation}
By requiring the relativistic invariance of the variational
principle in Eq. (\ref{4.5a}), Majorana deduces both the
transformation law for $\psi$ under an (infinitesimal) Lorentz
transformation and the explicit expressions for the matrices
${\bvec{\alpha}}$, $\beta$. In particular, the transformation law
for $\psi$ is obtained directly from the corresponding ones for
the variables $ct, x, y, z$ by means of the matrices $a$ and $b$
in the representation (\ref{4.5}). By using the same procedure
leading to the matrix elements in (\ref{4.4b}), Majorana gets the
following expressions for the elements of the (infinite) Dirac
$\bvec{\alpha}$ and $\beta$ matrices:
 \bea
\!\!\!\!\! \dps <j,m \, | \, \alpha_x - i \alpha_y \, | \,
j+1,m+1> &=& \dps - \,  1/2  \, \sqrt{\frac{(j+m+1)(j+m+2)}{\dps
\left( j+ 1/2  \right) \left( j +  3/2  \right)}},
\nonumber \\
\!\!\!\!\! \dps <j,m \, | \, \alpha_x - i \alpha_y \, | \,
j-1,m+1> &=& \dps - \,  1/2  \, \sqrt{\frac{(j-m)(j-m-1)}{\dps
\left( j- 1/2  \right) \left( j +  1/2  \right)}},
\nonumber \\
\!\!\!\!\! \dps <j,m \, | \, \alpha_x + i \alpha_y \, | \,
j+1,m-1> &=& \dps  1/2  \, \sqrt{\frac{(j-m+1)(j-m+2)}{\dps \left(
j+ 1/2  \right) \left( j +  3/2  \right)}},
\nonumber \\
\!\!\!\!\! \dps <j,m \, | \, \alpha_x + i \alpha_y \, | \,
j-1,m-1> &=& \dps   1/2  \, \sqrt{\frac{(j+m)(j+m-1)}{\dps \left(
j- 1/2  \right) \left( j +  1/2  \right)}},
\label{4.5b}  \\
\!\!\!\!\! \dps <j,m \, | \, \alpha_z \, | \, j+1,m> &=& \dps - \,
 1/2  \, \sqrt{\frac{(j+m+1)(j-m+1)}{\dps \left(
j+ 1/2  \right) \left( j +  3/2  \right)}},
\nonumber \\
\!\!\!\!\! \dps <j,m \, | \, \alpha_z \, | \, j-1,m> &=& \dps - \,
 1/2  \, \sqrt{\frac{(j+m)(j-m)}{\dps \left( j- 1/2
\right) \left( j +  1/2  \right)}}, \nonumber \\
\!\!\!\!\! \dps \beta &=& \frac{1}{\dps j +  1/2 } . \nonumber
 \eea
The Majorana equation for particles with arbitrary spin has, then,
the same form of the Dirac equation:
\begin{equation}\label{4.6}
  \left( \frac{H}{c} + {\bvec{\alpha}}\cdot {\bvec{p}} - \beta m c
  \right) \psi \, = \, 0 ,
\end{equation}
but with different (and infinite) matrices $\alpha$ and $\beta$,
whose elements are given in Eqs. (\ref{4.5b}). The rest energy of
the particles thus described has the form:
\begin{equation}\label{4.7}
  E_0 \, = \, \frac{m c^2}{\displaystyle s + 1/2} ,
\end{equation}
and depends on the spin $s$ of the particle. We here stress that
the scientific community of that time was convinced that only
equations of motion for spin 0 (Klein-Gordon equation) and spin
1/2 (Dirac equation) particles could be written down. The
importance of the Majorana work was first realized by van der
Waerden \cite{vdw} but, unfortunately, the paper remained
unnoticed until recent times.

\vspace{2cm} \noindent {\Large \bf Acknowledgments}

\vspace{1truecm}

\noindent The author warmly thanks I. Licata for his kind spur to
write the present paper and E. Recami for fruitful discussions.

\end{document}